\documentclass{article}
\usepackage{amssymb}

\usepackage{amsmath}

\input{tcilatex}

\begin{document}

\title{A Pragmatic Interpretation of Quantum Logic}
\author{Claudio Garola \\
Dipartimento di Fisica, Universit\`{a} di Lecce e Sezione INFN\\
73100 Lecce, Italy\\
E-mail: Garola@le.infn.it}
\maketitle

\begin{abstract}
Scholars have wondered for a long time whether quantum mechanics (QM)
subtends a quantum concept of truth which originates quantum logic (QL) and
is radically different from the classical (Tarskian) concept of truth. We
show in this paper that QL can be interpreted as a pragmatic language $%
\mathcal{L}_{QD}^{P}$ of pragmatically decidable assertive formulas, which
formalize statements about physical systems that are empirically \textit{%
justified} or \textit{unjustified} in the framework of QM. According to this
interpretation, QL formalizes properties of the metalinguistic concept of
empirical justification within QM rather than properties of a quantum
concept of truth. This conclusion agrees with a general integrationist
perspective, according to which nonstandard logics can be interpreted as
theories of metalinguistic concepts different from truth, avoiding
competition with classical notions and preserving the globality of logic. By
the way, some elucidations of the standard concept of quantum truth are also
obtained.\medskip

\end{abstract}

\section{Introduction}

The formal structure called \textit{quantum logic} (QL) springs out in a
natural way from the formalism of quantum mechanics (QM). Scholars have
debated for a long time on it, wondering whether it subtends a concept of
quantum truth which is typical of QM, and a huge literature exists on this
topic. We limit ourselves here to quote the classical book by Jammer,$^{(1)}$
which provides a general review of QL from its birth to the early seventies,
and the recent books by R\`{e}dei$^{(2)}$ and Dalla Chiara \textit{et al.},$%
^{(3)}$ which contain updated bibliographies.

Whenever the existence of a quantum concept of truth is accepted, one sees
at once that it has to be radically different from the classical (Tarskian)
concept, since the set of propositions of QL has an algebraic structure
which is different from the structure of classical propositional logic.
Thus, a new problem arises, \textit{i.e}. the problem of the ``correct''
logic to be adopted when reasoning in QM.

We want to show in the present paper that the above problem can be avoided
by adopting an \textit{integrated perspective}, which preserves both the
globality of logic (in the sense of \textit{global pluralism}, which admits
the existence of a plurality of mutually compatible logical systems, but not
of systems which are mutually incompatible$^{(4)}$) and the classical notion
of \textit{truth as correspondence}, which we consider as explicated
rigorously by Tarski's semantic theory.$^{(5,6)}$ This perspective
reconciliates non-Tarskian theories of truth with Tarski's theory by
reinterpreting them as theories of metalinguistic concepts that are
different from truth, and can be fruitfully applied to QL. Indeed, we prove
in this paper that QL can be interpreted as a theory of the concept of 
\textit{empirical justification} within QM.

In order to grasp intuitively our results, let us anticipate briefly some
remarks that will be discussed more extensively in Sec. 2.

First of all, it must be noted that QM usually avoids making statements
about properties of individual samples of a physical system (\textit{%
physical objects}). Rather, it is concerned with probabilities of results of
measurements on physical objects (\textit{standard interpretation}, as
espounded in any manual of QM; see, \textit{e.g}., Refs. 7, 8 and 9), or
with statistical predictions about ensembles of identically prepared
physical objects (\textit{statistical interpretation}; see, \textit{e.g}.,
Refs. 1, 10 and 11). Yet, QM also distinguishes between properties that are
real (or \textit{actual}) and properties that are not real (or \textit{%
potential}) in a given state $S$ of the physical system that is considered
(briefly, the property $E$ is actual in $S$ whenever a test of $E$ on any
physical object $x$ in $S$ would show that $E$ is possessed by $x$ without
changing $S^{(12)}$). This amounts to introduce implicitly a concept of
truth that also applies to statements about individuals. Indeed, asserting
that a property $E$ is actual in the state $S$ is equivalent to asserting
that the statement $E(x)$ that attributes $E$ to a physical object $x$ is 
\textit{true} for every $x$ in the state $S$. Moreover, according to QM, $%
E(x)$ is true, for a given $x$ in the state $S$, if and only if (briefly, 
\textit{iff}) $E$ is actual in the state $S$.$^{(12)}$ Falsity is then
defined by considering a complementary property $E^{\bot }$ of $E$, so that $%
E(x)$ is false for a given $x$ in the state $S$ iff $E^{\bot }$ is actual in 
$S$. It follows in particular that $E(x)$ is true (false) for a given $x$%
\textit{\ }in the state\textit{\ }$S$ iff it is true (false) for every $x$
in $S$, or, equivalently, iff it is \textit{certainly true} (\textit{%
certainly false}) in $S$. This result explains the notion of true as \textit{%
certain} introduced in some well known approaches to QM$^{(13,14)}$. More
important, it shows that the notion of truth has very peculiar features in
QM. Indeed, the truth and falsity of a statement $E(x)$ about an individual
are equivalent to the truth of two universally quantified statements. Both
these statements may be false. In this case $E(x)$ has no truth value, hence
it is meaningless. The existence of meaningless statements implies, in
particular, that no Tarskian set-theoretical semantics can be introduced in
QM.

The quantum notion of truth and meaning pointed out above is typical of the
standard interpretation of QM, and it is inspired by a verificationist
position which identifies truth and verifiability, meaning and verifiability
conditions. These identifications are rather doubtful from an
epistemological viewpoint, yet it is commonly maintained in the literature
that the standard quantum conception of truth has no alternatives, since it
seems firmly rooted in the formalism of QM itself. The mathematical
apparatus of QM would imply indeed the impossibility of defining an
assignment function associating a truth value with every individual
statement of the form $E(x)$ by referring only to the property $E$ and the
state $S$ of $x$. The outcomes obtained in a concrete experiment whenever $E$
or $E^{\bot }$ are not actual in $S$ would depend on the set of observations
that are carried out simultaneously, not only on $S$ (\textit{contextuality}%
).$^{(15-18)}$

Notwithstanding the arguments supporting it, the standard viewpoint can be
criticized, and an alternative \textit{SR interpretation} of QM can be
constructed based on an epistemological position (\textit{semantic realism},
or, briefly \textit{SR}) which allows one to define a truth value for every
statement of the form $E(x)$ according to a Tarskian set-theoretical model.$%
^{(19-26)}$ Of course, all statements that are certainly true (equivalently,
true) or certainly false (equivalently, false) according to the standard
interpretation with its quantum concept of truth, are also certainly true or
certainly false, respectively, according to the SR interpretation with its
Tarskian concept of truth. The remaining statements are meaningless
according to the former interpretation, while they have truth values
according to the latter: these values, however, may change when different
objects in the same state are considered, and cannot be predicted in QM
(which is, in this sense, an incomplete theory).

Because of its intuitive, philosophical and technical advantages, we adopt
the SR interpretation in the present paper. It is then important to observe
that our definitions and reasonings take into account only statements that
are certainly true (certainly false) in the sense explained above, hence
they actually do not depend on the choice of the interpretation of QM
(standard or SR). Thus, our reinterpretation of QL should be acceptable also
for logicians and physicists who do not agree with our epistemological
position. Of course, if the SR interpretation is not accepted one loses all
philosophical advantages of the integrated perspective mentioned at the
beginning of this section.

Let us come now to empirical justification. Whenever a statement $E(x)$ is
certainly true (certainly false), its truth (falsity) can be predicted
within QM if the property $E$ and the state $S$ of $x$ are known, and can be
checked (by means of nontrivial physical procedures, see Sec. 2.6). Hence,
we can say that the assertion of $E(x)$ ($E^{\bot }(x)$) is empirically
justified, since we can both deduce the truth of $E(x)$ ($E^{\bot }(x)$)
inside QM and provide an empirical proof of it. More formally, one can
introduce an assertion sign $\vdash $ and say that $E(x)$ is certainly true
(certainly false) iff $\vdash E(x)$ ($\vdash E^{\bot }(x)$) is empirically
justified. In this way a semantic notion (certainty of truth) is translated
into a pragmatic notion (empirical justification). Now, we remind that a
pragmatic extension of a classical language and some general properties of
the concept of justification have been studied by Dalla Pozza and by the
author$^{(27)}$ and note that all results obtained in this research apply to
the notion of empirical justification introduced above. Moreover, further
results can be obtained which are typical of the case under consideration,
since the notion of justification is now specified (empirical justification
in QM). Thus, a pragmatic language $\mathcal{L}_{Q}^{P}$ can be constructed
(Sec. 3) in which assertions of the form $\vdash E(x)$ are taken as
elementary \textit{assertive formulas} (\textit{afs}) and pragmatic
connectives are introduced, for which a \textit{set-theoretical pragmatics}
is defined basing on the concept of empirical justification in QM. This
pragmatics defines a justification value for every elementary or complex af
of $\mathcal{L}_{Q}^{P}$, yet not all complex afs of $\mathcal{L}_{Q}^{P}$
are \textit{pragmatically decidable}, that is, such that an empirical
procedure of justification exists (it obviously exists for all elementary
afs of $\mathcal{L}_{Q}^{P}$ because of our arguments above). However, one
can single out a subset of pragmatically decidable afs of $\mathcal{L}%
_{Q}^{P}$ and consider a sublanguage $\mathcal{L}_{QD}^{P}$ of $\mathcal{L}%
_{Q}^{P}$ which contains only afs in this subset. It is then easy to see
that our set-theoretical pragmatics, when restricted to $\mathcal{L}%
_{QD}^{P} $, endows it with the structure of QL.

The above result is highly interesting in our opinion. Indeed, it provides
the desired reinterpretation of QL as a theory of the metalinguistic concept
of empirical justification in QM, allowing us to place it within an
integrationist perspective that avoids any conflict with classical logic (we
stress again that this conclusion can be accepted also by scholars who want
to maintain the standard interpretation of QM).

We conclude this Introduction by observing that our results suggest that the
standard partition of properties in two subsets (actual properties and
potential properties) should be substituted by a partition in three subsets,
as follows.

\textit{Actual properties}. A property $E$ is actual in the state $S$ iff
the assertion $\vdash E(x)$, with $x$ in $S$, is justified.

\textit{Nonactual properties}. A property $E$ is nonactual in the state $S$
iff the assertion $\vdash E^{\bot }(x)$, with $x$ in $S$, is justified.

\textit{Potential properties}. A property $E$ is potential in the state $S$
iff both assertions $\vdash E(x)$ and $\vdash E^{\bot }(x)$, with $x$ in $S$%
, are unjustified.

\section{Physical preliminaries}

We introduce in this section a number of symbols, definitions and physical
concepts that will be extensively used in Sec. 3 in order to supply an
intuitive support and an intended interpretation for the pragmatic language
that will be introduced there.

\subsection{Basic notions and mathematical representations}

The following notions will be taken as primitive.

\textit{Physical system }$\Omega $\textit{.}

\textit{Pure state }$S$\textit{\ of }$\Omega $, and \textit{set $\mathcal{S}$
of all pure states of }$\Omega $ (the word \textit{pure} will be usually
implied in the following).

\textit{Testable property }$E$\textit{\ of }$\Omega $, and \textit{set $%
\mathcal{E}$ of all testable properties of }$\Omega $ (the word \textit{%
testable} will be usually implied in the following).\footnote{%
It must be noted that the physical properties considered here are first
order properties from a logical viewpoint.$^{(26)}$ Higher order properties
obviously occur in physics and will be encountered later on (Sec. 2.6), but
they need not be considered here.}

States and properties will be interpreted operationally as follows.

A state $S\in \mathcal{S}$ is a class of physically equivalent\footnote{%
The notion of physical equivalence for preparing or registering devices is
not trivial.$^{(11,21)}$ We do not discuss it here for the sake of brevity.}
preparing devices (briefly, \textit{preparations}) which may prepare
individual samples of $\Omega $ (\textit{physical objects}). A physical
object $x$ \textit{is in the state }$S$ iff it is prepared by a preparation $%
\pi \in S$.

A property $E\in \mathcal{E}$ is a class of physically equivalent ideal
dichotomic (outcomes 1, 0) registering devices (briefly, \textit{%
registrations}) which may test physical objects.\footnote{%
Note that a registration may act as a new preparation of the physical object 
$x$, so that the state of $x$ may change after a test on it.}

The above notions do not distinguish between classical and quantum
mechanics. The mathematical representation of physical systems, states and
properties are different, however, in the two theories. Let us resume these
representations in the case of QM.

Every physical system $\Omega $ is associated with a Hilbert space $\mathcal{%
H}$ over the field of complex numbers (we use the Dirac notation $\mid \cdot
\rangle $ in order to denote vectors of $\mathcal{H}$ in the following).

Let us denote by $(\mathcal{L(H)},\subset )$ the partially ordered set
(briefly, \textit{poset}) of all closed subspaces of $\mathcal{H}$ (here $%
\subset $ denotes set-theoretical inclusion), and let $\mathcal{A}\subset 
\mathcal{L(H)}$ be the set of all one-dimensional subspaces of $\mathcal{H}$%
. Then (in absence of superselection rules) a mapping

\smallskip

$\varphi :S\in \mathcal{S\longrightarrow \varphi }(S)\in \mathcal{A}$

\smallskip

exists which maps bijectively the set $\mathcal{S}$ of all states of $\Omega 
$ onto $\mathcal{A}$,\footnote{%
It follows easily that every state S can also be represented by any vector $%
\mid \psi \rangle \in \varphi (S)\in \mathcal{A}$, which is the standard
representation adopted in elementary QM. Moreover, a state S is usually
represented by an (orthogonal) projection operator on $\varphi (S)$ in more
advanced QM. However, the representation $\varphi $ introduced here is more
suitable for our purposes in the present paper.} and a mapping

\smallskip

$\chi :E\in \mathcal{E\longrightarrow \chi }(E)\in \mathcal{L(H)}$

\smallskip

exists which maps bijectively the set $\mathcal{E}$ of all properties of $%
\Omega $ onto $\mathcal{L(H)}$.\footnote{%
Equivalently, a property is often represented in QM as a pair $(A,\Delta )$,
where is $A$ a self-adjoint operator on $\mathcal{H}$ representing a
physical observable, and $\Delta $ a Borel set on the real line.$^{(28)}$ We
do not use this representation, however, in the present paper.}

\subsection{Physical Quantum Logic}

The poset $(\mathcal{L(H)},\subset )$ is characterized by a set of
mathematical properties. In particular, it is a complete, orthocomplemented,
weakly modular, atomic lattice which satisfies the covering law$%
^{(13,27-30)} $. We denote by $^{\bot }$, $\Cap $ and $\Cup $
orthocomplementation, meet and join, respectively, in $(\mathcal{L(H)}%
,\subset )$, and remind that $\Cap $ coincides with the set-theoretical
intersection $\cap $ of subspaces of $\mathcal{H}$, while $^{\bot }$ does
not generally coincide with the set-theoretical complementation $^{\prime }$%
, nor $\Cup $ coincides with the set-theoretical union $\cup $. Furthermore,
we denote the minimal element $\{\mid 0\rangle \}$ and the maximal element $%
\mathcal{H}$ of $(\mathcal{L(H)},\subset )$ by $O$ and $I$, respectively.
Finally, we note that $\mathcal{A}$ obviously coincides with the set of all
atoms of $(\mathcal{L(H)},\subset )$.

Let us denote by $\prec $ the order induced on $\mathcal{E}$, via the
bijective representation $\chi $, by the order $\subset $ defined on $%
\mathcal{L(H)}$. Then, the poset $(\mathcal{E},\prec )$ is order-isomorphic
to $(\mathcal{L(H)},\subset )$, hence it is characterized by the same
mathematical properties characterizing $(\mathcal{L(H)},\subset )$. In
particular, the unary operation induced on it, via $\chi $, by the
orthocomplementation defined on $(\mathcal{L(H)},\subset )$, is an
orthocomplementation, and $(\mathcal{E},\prec )$ is an orthomodular (i.e.,
orthocomplemented and weakly modular) lattice, usually called \textit{the
lattice of properties} of $\Omega $. By abuse of language, we denote the
lattice operations on $(\mathcal{E},\prec )$ by the same symbols used above
in order to denote the corresponding lattice operations on $(\mathcal{L(H)}%
,\subset )$.

Orthomodular lattices are said to characterize semantically \textit{%
orthomodular QLs }in the literature.$^{(3)}$ The lattice of properties has a
less general structure in QM, since it inherits a number of further
properties from $(\mathcal{L(H)},\subset )$. Therefore, $(\mathcal{E},\prec
) $ will be called \textit{physical QL} in this paper.

A further lattice, isomorphic to $(\mathcal{E},\prec )$, will be used in the
following. In order to introduce it, let us consider the mapping

\smallskip

$\rho :E\in \mathcal{E}\longrightarrow \mathcal{S}_{E}=\{S\in \mathcal{S}%
\mid \varphi (S)\subset \chi (E)\}\in \mathcal{L(S)}$,

\smallskip

where $\mathcal{L(S)}=\{\mathcal{S}_{E}\mid E\in \mathcal{E}\}$ is the range
of $\rho $, which generally is a proper subset of the power set $\mathcal{%
P(S)}$ of $\mathcal{S}$. The poset $(\mathcal{L(S)},\subset )$ is
order-isomorphic to $(\mathcal{L(H)},\subset )$, hence to $(\mathcal{E}%
,\prec )$, since $\varphi $ and $\chi $ are bijective, so that $\rho $ is
bijective and order-preserving. Therefore $(\mathcal{L(S)},\subset )$ is
characterized by the same mathematical properties characterizing $(\mathcal{E%
},\prec )$. In particular, the unary operation induced on it, via $\rho $,
by the orthocomplementation defined on $(\mathcal{E},\prec )$, is an
orthocomplementation, and $(\mathcal{L(S)},\subset )$ is an orthomodular
lattice. We denote orthocomplementation, meet and join on $(\mathcal{L(S)}%
,\subset )$ by the same symbols $^{\bot }$, $\Cap $, and $\Cup $,
respectively, that we have used in order to denote the corresponding
operations on $(\mathcal{L(H)},\subset )$ and $(\mathcal{E},\prec )$, and
call $(\mathcal{L(S)},\subset )$ \textit{the lattice of closed subsets of }$%
\mathcal{S}$ (the word \textit{closed} refers here to the fact that, for
every $\mathcal{S}_{E}\in $ $\mathcal{L(S)}$, $(\mathcal{S}_{E}^{\bot
})^{\bot }=\mathcal{S}_{E}$). We also note that the operation $\Cap $
coincides with the set-theoretical intersection $\cap $ on $\mathcal{L(S)}$
because of the analogous result holding in $(\mathcal{L(H)},\subset )$.%
\footnote{%
Whenever the dimension of $\mathcal{H}$ is finite, the lattice $(\mathcal{%
L(H)},\subset )$ and/or the lattice $(\mathcal{L(S)},\subset )$ can be
identified with Birkhoff and von Neumann's lattice of \textit{experimental
propositions}, which was introduced in the 1936 paper that started the
research on QL.$^{(31)}$ This identification is impossible, however, if $%
\mathcal{H}$ is not finite-dimensional, since Birkhoff and von Neumann's
lattice is modular, not simply weakly modular. The requirement of modularity
has deep roots in the von Neumann concept of probability in QM according to
some authors.$^{(2)}$}

To close up, let us observe that the unary operation $^{\bot }$ defined on $%
\mathcal{L(S)}$ can be extended to $\mathcal{P(S)}$ by setting, for every $%
\mathcal{T\in P(S)}$,

$\mathcal{T}^{\bot }=($\textit{min}$\{\mathcal{S}_{E}\in \mathcal{L(S)\mid
T\subset S}_{E}\})^{\bot }$

(the symbol \textit{min} obviously refers to the order $\subset $ defined on 
$\mathcal{L(S)}$). This extension will be needed indeed in Sec. 3.2.

\subsection{Actual and potential properties}

We say that a property $E$ is\textit{\ actual} (\textit{nonactual}) in the
state $S$ iff one can perform a test of $E$ on any physical object $x$ in
the state $S$ by means of a registration $r\in E$, obtaining outcome 1 (0)
without modifying $S$.\footnote{%
One can provide an intuitive support to this definition by noticing that the
result obtained in a test of $E$ on a physical object $x$ in the state $S$
can be attributed to $x$ only whenever $S$ is not modified by the test.
Moreover, only in this case the test is \textit{repeatable}, i.e., it can be
performed again obtaining the same result.
\par
It is well known that classical physics assumes that tests which do not
modify the state $S$ are always possible, at least as ideal limits of
concrete procedures, while this assumption does not hold in QM.}

Basing on the above definition, for every state $S\in \mathcal{S}$ three
subsets of $\mathcal{E}$ can be introduced.

$\mathcal{E}_{S}$ : the set of all properties that are actual in $S$.

$\mathcal{E}_{S}^{\bot }$ : the set of all properties that are nonactual in $%
S$.

$\mathcal{E}_{S}^{I}$ : the set $\mathcal{E}\setminus \mathcal{E}_{S}\cup 
\mathcal{E}_{S}^{\bot }$ (called the set of all properties that are \textit{%
indeterminate}, or\textit{\ potential}, in $S$).

By using the mathematical apparatus of QM, the sets $\mathcal{E}_{S}$ and $%
\mathcal{E}_{S}^{\bot }$ can be characterized as follows.

\smallskip

$\mathcal{E}_{S}=\{E\in \mathcal{E}\mid \varphi (S)\subset \chi (E)\}=\{E\in 
\mathcal{E}\mid S\in \mathcal{S}_{E}\}$.

\smallskip

$\mathcal{E}_{S}^{\bot }=\{E\in \mathcal{E}\mid \varphi (S)\subset \chi
(E)^{\bot }\}=\{E\in \mathcal{E}\mid S\in \mathcal{S}_{E}^{\bot }\}$.

\smallskip

It can also be proved that $\mathcal{E}_{S}$ ($\mathcal{E}_{S}^{\bot }$)
coincides with the set of all properties that have probability 1 (0),
according to QM, for every $x$ in the state $S$, and that the set $\mathcal{E%
}_{S}^{I}$ (which is non-void in QM, while it would be void in classical
physics) coincides with the set of all properties that have probability
different from 0 and 1 for every $x$ in the state $S$.

Further characterizations of the above sets can be obtained as follows.$%
^{(12)}$

Since the mapping $\rho $ is bijective, while every singleton $\{S\}$, with $%
S\in \mathcal{S}$, obviously is an atom of $\mathcal{L(S)}$, one can
associate a property $E_{S}=\rho ^{-1}(\{S\})$ (equivalently, $E_{S}=\chi
^{-1}(\varphi (S))$) with every $S\in \mathcal{S}$. This property is an atom
of $(\mathcal{E},\prec )$, and is usually called the \textit{support} of $S$%
. The mapping $\rho ^{-1}$ thus induces a one-to-one correspondence between
(pure) states and atoms of $(\mathcal{E},\prec )$. Then, one can prove the
following equalities.

\smallskip

$\mathcal{E}_{S}=\{E\in \mathcal{E}\mid E_{S}\prec E\}$.

\smallskip

$\mathcal{E}_{S}^{\bot }=\{E\in \mathcal{E}\mid E\prec E_{S}^{\bot }\}$.

\smallskip

$\mathcal{E}_{S}^{I}=\{E\in \mathcal{E}\mid E_{S}\nprec E$ and $E\nprec
E_{S}^{\bot }\}$.

\smallskip

Finally, the following equality also follows from the above definitions.

\smallskip

$\mathcal{S}_{E}=\{S\in \mathcal{S}\mid E_{S}\prec E\}.$

\subsection{Truth in standard QM}

No mention has been done of truth values (\textit{true/false}) in the
foregoing sections. However, we will be concerned with logical structures in
Sec. 3, hence it is natural to wonder what QM says about the truth of a
sentence as ``the physical object $x$ has the property $E$'' (briefly, $E(x)$
in the following).

We have already noted in the Introduction that QM usually avoids making
explicit statements regarding individual samples of physical systems. Yet, a
sentence as ``the property $E$ is actual in the state $S$'' (Sec. 2.3)
intuitively means that all physical objects in the state $S$ have the
property $E$. Hence, it can be translated, in terms of truth, into the
sentence ``for every physical object $x$ in the state $S$, $E(x)$ is true''.
This translation shows that QM is concerned also with truth values of
individual statements. Moreover, by considering the literature on the
subject, one can argue that QM more or less implicitly adopts the following
verificationist criterion of truth.$^{(12)}$

\smallskip

EV (empirical verificationism). \textit{The sentence }$E(x)$\textit{\ has
truth value} true\textit{\ (}false\textit{) for a physical object }$x$%
\textit{\ in the state }$S$\textit{\ iff }$E$\textit{\ is actual (nonactual)
in }$S$\textit{, while it is meaningless otherwise.}

\smallskip

Criterion EV is obviously at odds with standard definitions in classical
logic (CL), and is suggested by the fact that $E$ can be attributed (not
attributed) to a physical object $x$ in the state S on the basis of an
experimental procedure only when it is actual (nonactual) for $x$ (see Sec.
2.6). Hence, we say that $E(x)$ is \textit{Q-true} (\textit{Q-false})
whenever its truth value is true (false) according to criterion EV, in order
to stress the difference between the truth values introduced in QM and those
introduced in CL.

Because of the foregoing translation, criterion EV implies the following
proposition.

\smallskip

TF. \textit{The sentence }$E(x)$\textit{\ is }Q-true\textit{\ (}Q-false%
\textit{) for a physical object x in the state }$S$\textit{\ iff it is }%
Q-true\textit{\ (}Q-false\textit{) for every physical object x in the state }%
$S$\textit{.}

\smallskip

Loosely speaking, proposition TF can be rephrased by saying that $E(x)$ is
true (false) in the sense established by criterion EV iff it is \textit{%
certainly }true (\textit{certainly }false) in the same sense, which explains
the intuitive terminology that we have adopted in the Introduction.

Furthermore, criterion EV implies that $E(x)$ has a truth value in standard
QM iff $E\in \mathcal{E}_{S}\cup \mathcal{E}_{S}^{\bot }$ (of course, $E(x)$
is Q-true iff $E\in \mathcal{E}_{S}$, Q-false iff $E\in \mathcal{E}%
_{S}^{\bot }$). It is then important to observe that the characterizations
of $\mathcal{E}_{S}$ and $\mathcal{E}_{S}^{\bot }$ provided in Sec. 2.3 show
that, for every $S\in \mathcal{S}$, one can deduce from theoretical laws of
QM whether a property $E$ belongs to $\mathcal{E}_{S}\cup \mathcal{E}%
_{S}^{\bot }$. In particular, $E$ belongs to $\mathcal{E}_{S}$ ($\mathcal{E}%
_{S}^{\bot }$) iff it has probability 1 (0) for every $x$ in the state $S$.
Hence, one can predict, for every $E\in \mathcal{E}$ and $x$ in the state $S$%
, whether $E(x)$ is Q-true, Q-false or meaningless. This result shows that
standard QM is a semantically complete theory$^{(12)}$ and, together with
proposition TF, explains the definition of \textit{true} as \textit{certain}%
, or \textit{predictable}, which occurs in some approaches to QM.$^{(13,14)}$

\subsection{Nonobjectivity versus objectivity in QM}

The position expounded in Sec. 2.4 about the truth value of sentences of the
form $E(x)$, with $E\in \mathcal{E}$, is sometimes summarized by saying,
briefly, that physical properties are \textit{nonobjective} in standard QM
(to be precise, only the properties in $\mathcal{E}_{S}^{I}$ should be
classified as nonobjective for a given $S\in \mathcal{S}$).

Nonobjectivity of properties is supported by a number of arguments. Some of
them are based on empirical results (e.g., the two-slits experiment), some
follow from seemingly reasonable epistemological choices (e.g., the adoption
of a verificationist position, together with the indeterminacy principle)
and some take the form of theorems deduced from the mathematical apparatus
of QM. These last arguments are usually considered conclusive in the
literature. We remind here the Bell-Kochen-Specker and Bell's theorems$%
^{(15-18)}$ which seem to prove that it is impossible to assign classical
truth values to all sentences of the form $E(x)$, with $E\in \mathcal{E}$,
without contradicting the predictions of QM.

However, all arguments which show that nonobjectivity of properties is an
unavoidable feature of QM can be criticized (this of course does not make
the claim of nonobjectivity wrong, but only proves that there are
alternatives to it). In particular, one can observe that a \textit{no-go
theorem} as Bell-Kochen-Specker's is certainly correct from a mathematical
viewpoint, but rests on implicit assumptions that are problematic from a
physical and epistemological viewpoint.$^{(22-25)}$ Basing on this
criticism, an alternative interpretation (\textit{semantic realism}, or 
\textit{SR}, interpretation) has been propounded by the author, together
with other authors.$^{(19-23,25,26)}$ As we have already observed in the
Introduction, the SR interpretation adopts a Tarskian theory of truth as
correspondence, and all properties are objective according to it
(equivalently, the sentence $E(x)$ has a truth value defined in a classical
way for every physical object $x$ and property $E$). According to this
interpretation $E(x)$ is \textit{certainly true} (\textit{certainly false})
in the state $S$, that is, it is true (false) in a classical sense for every 
$x$ is in the state $S$, iff $E\in \mathcal{E}_{S}$ ($E\in \mathcal{E}%
_{S}^{\bot }$), hence iff it is Q-true (Q-false) according to the standard
interpretation.

The SR interpretation of QM has some definite advantages. Firstly, it makes
QM compatible with a realistic perspective without requiring any change of
its mathematical apparatus and preserving all statistical predictions
following from the standard interpretation, hence it provides a solution of
the quantum measurement problem.$^{(26)}$ Secondly, it rests on a classical
conception of truth and meaning. Thirdly, it leads one to consider QM as an
incomplete theory,$^{(12)}$ and provides some suggestions about the way in
which a more general theory embodying QM could be constructed.

Also within the SR interpretation one can deduce from theoretical laws of QM
whether $E\in \mathcal{E}_{S}$ ($E\in \mathcal{E}_{S}^{\bot }$), for a given 
$S\in \mathcal{S}$. Moreover, for every $E\in \mathcal{E}_{S}\cup \mathcal{E}%
_{S}^{\bot }$, the sentence $E(x)$ obviously is certainly true, hence true
(certainly false, hence false) iff $E\in \mathcal{E}_{S}$ ($E\in \mathcal{E}%
_{S}^{\bot }$). On the contrary, no prediction of the truth value of $E(x)$
can be done if $E\notin \mathcal{E}_{S}\cup \mathcal{E}_{S}^{\bot }$. Thus,
the difference between the standard and the SR interpretation reduces to the
fact that, whenever $E\in \mathcal{E}_{S}^{I}$, $E(x)$ is meaningless within
the former, while it has a truth value that cannot be predicted by QM within
the latter.

\subsection{Empirical proof in QM}

The results at the end of Secs. 2.4 and 2.5 show that, whenever $x$ is in
the state $S$, the truth value of the sentence $E(x)$ can be predicted (or 
\textit{theoretically proved}) iff $E\in \mathcal{E}_{S}\cup \mathcal{E}%
_{S}^{\bot }$, both in the standard and in the SR interpretation. One is
thus led to wonder whether and when the truth value of $E(x)$ can be
determined empirically. At first glance, it seems sufficient to test $x$ by
means of a registering device belonging to $E$ (Sec. 2.1). This is untrue
according to the standard as well as the SR interpretation. Indeed, both
interpretations maintain that a single test modifies, whenever $E\notin 
\mathcal{E}_{S}\cup \mathcal{E}_{S}^{\bot }$, the state $S$ of the physical
object $x$, so that its result refers to the final state after the test,
which is different from $S$ (moreover, within the standard interpretation, $%
E(x)$ has no truth value whenever $E\notin \mathcal{E}_{S}\cup \mathcal{E}%
_{S}^{\bot }$). Thus, a test of $E(x)$ is physically meaningful iff $E\in 
\mathcal{E}_{S}\cup \mathcal{E}_{S}^{\bot }$, since only in this case it
does not modify the state $S$. It follows that an \textit{empirical proof}
of the truth value of $E(x)$ can be given iff a theoretical proof of this
value exists, and it consists in checking whether $E\in \mathcal{E}_{S}$ or $%
E\in \mathcal{E}_{S}^{\bot }$. Then, the characterizations of $\mathcal{E}%
_{S}$ and $\mathcal{E}_{S}^{\bot }$ in Sec. 2.3 suggest the empirical
procedures to be adopted. Indeed, they show that $E\in \mathcal{E}_{S}$ ($%
E\in \mathcal{E}_{S}^{\bot }$) iff $E_{S}\prec E$ ($E\prec E_{S}^{\bot }$),
or, equivalently, iff $S\in \mathcal{S}_{E}$ ($S\in \mathcal{S}_{E}^{\bot }$%
). Hence, one can get an empirical proof that $E(x)$ is Q-true (Q-false)
within the standard interpretation, or equivalently, that $E(x)$ is
certainly true, hence true (certainly false, hence false) within the SR
interpretation, by checking whether the state $S$ of $x$ belongs to the set $%
\mathcal{S}_{E}$ ($\mathcal{S}_{E}^{\bot }$). The empirical procedure
required by this check is rather complex, since it does not reduce to a test
of $E$ on the physical object $x$, but consists in testing a huge number of
physical objects in the state $S$ by means of registrations belonging to $E$%
, in order to show that all of them yield outcome 1 (0) (it has been proven
elsewhere$^{(26)}$ that this procedure actually tests a quantified
statement, or a second order physical property).

We conclude by noticing that truth and empirical provability of truth
coincide within the standard interpretation of QM, which expresses the
verificationist position that characterizes this interpretation. On the
contrary, within the SR interpretation of QM the concepts of truth and
empirical provability of truth are different, in accordance with the well
known distinction between truth and epistemic accessibility of truth in
classical logic.

\section{QL as a pragmatic language}

We aim to show in this section that physical QL can be recovered as a
pragmatic language in the sense established in Ref. 27. It is noteworthy
that, by weakening slightly the assumptions introduced in Ref. 27, one could
perform this task without choosing between the standard and the SR
interpretation of QM (see footnotes 8 and 9). We adopt however the SR
interpretation in this section, since we maintain that the verificationist
attitude of the standard interpretation is epistemologically and
philosophically doubtful, but we point out by means of footnotes the simple
changes to be introduced in order to attain the same results within the
standard interpretation.

\subsection{The general pragmatic language $\mathcal{L}^{P}$}

Let us summarize briefly the construction of the general pragmatic language $%
\mathcal{L}^{P}$ introduced in Ref. 27.

The alphabet $\mathcal{A}^{P}$ of $\mathcal{L}^{P}$ contains as \textit{%
descriptive signs }the propositional letters $p$, $q$, $r$,...; as\textit{\
logical-semantic signs} the connectives $\urcorner $, $\wedge $, $\vee $, $%
\rightarrow $ and $\leftrightarrow $; as \textit{\ logical-pragmatic signs}
the assertion sign $\vdash $ and the connectives $N$, $K$, $A$, $C$ and%
\textit{\ }$E$; as\textit{\ auxiliary signs} the round brackets $(.)$.%
\textit{\ }The set $\psi _{R}$ of all \textit{radical formulas} (\textit{rfs}%
) of $\mathcal{L}^{P}$ is made up by all formulas constructed by means of
descriptive and logical-semantic signs, following the standard recursive
rules of classical propositional logic (a rf consisting of a propositional
letter only is then called \textit{atomic}). The set $\psi _{A}$ of all 
\textit{assertive formulas }(\textit{afs}) of $\mathcal{L}^{P}$ is made up
by all rfs preceded by the assertive sign $\vdash $ (\textit{elementary}
afs), plus all formulas constructed by using elementary afs and following
standard recursive rules in which $N$, $K$, $A$, $C$ and\ $E$ take the place
of $\urcorner $, $\wedge $, $\vee $, $\rightarrow $ and $\leftrightarrow $,
respectively.

A \textit{semantic interpretation} of $\mathcal{L}^{P}$ is then defined as a
pair $(\{1,0\},\sigma )$, where $\sigma $ is an\textit{\ assignment function}
which maps $\psi _{R}$ onto the set $\{1,0\}$ of\textit{\ truth values} (1
standing for \textit{true} and 0 for \textit{false}), following the standard
truth rules of classical propositional calculus.

Whenever a semantic interpretation $\sigma $ is given, a \textit{pragmatic
interpretation} of $\mathcal{L}^{P}$ is defined as a pair $(\{J,U\},\pi
_{\sigma })$, where $\pi _{\sigma }$ is a \textit{pragmatic evaluation
function} which maps $\psi _{A}$ onto the set $\{J,U\}$ of \textit{\
justification values} following \textit{\ justification rules} which refer
to $\sigma $ and are based on the informal properties of the metalinguistic
concept of proof in natural languages. In particular, the following
justification rules hold.

\smallskip

JR$_{1}$. \textit{Let }$\alpha \in \psi _{R}$\textit{; then, }$\pi _{\sigma
}(\vdash \alpha )=J$\textit{\ iff a proof exists that }$\alpha $\textit{\ is
true, i.e., that }$\sigma (\alpha )=1$\textit{\ (hence, }$\pi _{\sigma
}(\vdash \alpha )=U$\textit{\ iff no proof exists that }$\alpha $\textit{\
is true).}

\smallskip

JR$_{2}$.\textit{\ Let }$\delta \in \psi _{A}$\textit{; then, }$\pi _{\sigma
}(N\delta )=J$\textit{\ iff a proof exists that }$\delta $\textit{\ is
unjustified, i.e., that }$\pi _{\sigma }(\delta )=U$\textit{.}

\smallskip

JR$_{3}$.\textit{\ Let }$\delta _{1}$\textit{, }$\delta _{2}\in \psi _{A}$%
\textit{; then,}

\textit{(i) }$\pi _{\sigma }(\delta _{1}K\delta _{2})=J$\textit{\ iff }$\pi
_{\sigma }(\delta _{1})=J$\textit{\ and }$\pi _{\sigma }(\delta _{2})=J$%
\textit{,}

\textit{(ii) }$\pi _{\sigma }(\delta _{1}A\delta _{2})=J$\textit{\ iff }$\pi
_{\sigma }(\delta _{1})=J$\textit{\ or }$\pi _{\sigma }(\delta _{2})=J$%
\textit{,}

\textit{(iii) }$\pi _{\sigma }(\delta _{1}C\delta _{2})=J$\textit{\ iff a
proof exists that }$\pi _{\sigma }(\delta _{2})=J$\textit{\ whenever }$\pi
_{\sigma }(\delta _{1})=J$\textit{,}

\textit{(iv) }$\pi _{\sigma }(\delta _{1}E\delta _{2})=J$\textit{\ iff }$\pi
_{\sigma }(\delta _{1}C\delta _{2})=J$\textit{\ and }$\pi _{\sigma }(\delta
_{2}C\delta _{1})=J$\textit{.}

\smallskip

Furthermore, the following \textit{correctness criterion} holds in $\mathcal{%
L}^{P}$.

\smallskip

CC. \textit{Let }$\alpha \in \psi _{R}$\textit{; then, }$\pi _{\sigma
}(\vdash \alpha )=J$\textit{\ implies }$\sigma (\alpha )=1.$

\smallskip

Finally, the set of all pragmatic evaluation functions that can be
associated with a given semantic interpretation $\sigma $ is denoted by $\Pi
_{\sigma }$.

\subsection{The quantum pragmatic language $\mathcal{L}_{Q}^{P}$}

The quantum pragmatic language $\mathcal{L}_{Q}^{P}$ that we want to
introduce here is obtained by specializing syntax, semantics and pragmatics
of $\mathcal{L}^{P}$. Let us begin with the syntax. We introduce the
following assumptions on $\mathcal{L}_{Q}^{P}$.

\smallskip

A$_{1}$. \textit{The propositional letters }$p$\textit{, }$q$\textit{, ...
are substituted by the symbols }$E(x)$\textit{, }$F(x)$\textit{, ..., with }$%
E$\textit{, }$F$\textit{, ... }$\in \mathcal{E}$\textit{.}

\smallskip

A$_{2}$. \textit{The set }$\psi _{R}^{Q}$\textit{\ of all rfs of }$\mathcal{L%
}_{Q}^{P}$\textit{\ reduces to the set of all atomic rfs of }$\mathcal{L}%
_{Q}^{P}$\textit{\ (in different words, if }$\alpha $\textit{\ is a rf of }$%
\mathcal{L}_{Q}^{P}$\textit{, then }$\alpha =E(x)$\textit{, with }$E\in 
\mathcal{E}$\textit{).}

\smallskip

A$_{3}$. \textit{Only the logical-pragmatic signs }$\vdash $\textit{, }$N$%
\textit{, }$K$\textit{\ and }$A$\textit{\ appear in the afs of }$\mathcal{L}%
_{Q}^{P}$\textit{.}

\smallskip

The substitution in A$_{1}$ aims to suggest the \textit{intended
interpretation} that we adopt in the following. To be precise, the rfs $E(x)$%
, $F(x)$, ... are interpreted as sentences stating that the physical object $%
x$ has the properties $E$, $F$, ..., respectively (Sec. 2.4).

The restriction in A$_{2}$ aims to select rfs that are interpreted as 
\textit{testable} sentences, i.e., sentences stating testable physical
properties (Sec. 2.1), so that physical procedures exist for testing their
truth values (which may not occur in the case of a rf of the form, say, $%
E(x)\vee F(x)$; note that a similar restriction has been introduced in Ref.
27 when recovering intuitionistic propositional logic within $\mathcal{L}%
^{P} $).

The restriction in A$_{3}$ is introduced for the sake of simplicity, since
only the pragmatic connectives $N$, $K$ and $A$ are relevant for our goals
in this paper.

Because of A$_{1}$, A$_{2}$ and A$_{3}$, the set $\psi _{A}^{Q}$ of afs of $%
\mathcal{L}_{Q}^{P}$ is made up by all formulas constructed by means of the
following recursive rules.

\smallskip

(i) \textit{Let }$E(x)$\textit{\ be a rf. Then }$\vdash E(x)$\textit{\ is an
af.}

(ii) \textit{Let }$\delta $\textit{\ be an af. Then, }$N\delta $\textit{\ is
an af.}

(iii) \textit{Let }$\delta _{1}$\textit{\ and }$\delta _{2}$\textit{\ be
afs. Then, }$\delta _{1}K\delta _{2}$\textit{\ and }$\delta _{1}A\delta _{2}$%
\textit{\ are afs.}

\smallskip

Let us come now to the semantics of $\mathcal{L}_{Q}^{P}$. We introduce the
following assumption on $\mathcal{L}_{Q}^{P}$.

\smallskip

A$_{4}$. \textit{Every assignment function }$\sigma $\textit{\ defined on }$%
\psi _{R}^{Q}$\textit{\ is induced by an interpretation }$\xi $\textit{\ of
the variable x that appears in the rfs into a universe }$\mathcal{U}$\textit{%
\ of physical objects, hence }$\sigma =\sigma (\xi )$\textit{\ and the
values of }$\sigma $ \textit{on }$\psi _{R}^{Q}$ \textit{are consistent with
(not necessarily determined by) the laws of QM within the intended
interpretation established above.}

\smallskip

Let us comment briefly on assumption A$_{4}$.

Firstly, note that the interpretation $\xi $ was understood in Sec. 2.1,
when we introduced the informal expression ``the physical object $x$ is in
the state $S$''.

Secondly, observe that the requirement that $\sigma =\sigma (\xi )$ be
consistent with the laws of QM (briefly, \textit{QM-consistent}) obviously
follows from the fact that these laws, via intended interpretation,
establish relations among the truth values of elementary rfs of $\mathcal{L}%
_{Q}^{P}$ whenever a specific physical object is considered. We denote by $%
\Sigma $ in the following the set of all QM-consistent assigment functions.

Thirdly, note that, since $\sigma =$ $\sigma (\xi )$, there may be many
interpretations of the variable x that lead to the same assigment function.

Finally, observe that the universe $\mathcal{U}$ can be partitioned into
(disjoint) subsets of physical objects, each of which consists of physical
objects in the same state (different subsets corresponding to different
states). Thus, specifying the state $S$ of $x$ means requiring that the
interpretation $\xi $ of $x$ that is considered maps $x$ on a physical
object in the subset corresponding to the state $S$, hence it singles out a
subclass $\Sigma _{S}\subset \Sigma $ of assigment functions. All functions
in $\Sigma _{S}$ assign truth value 1 (0) to a sentence $E(x)\in \psi
_{R}^{Q}$ whenever $E\in \mathcal{E}_{S}$ ($\mathcal{E}_{S}^{\bot }$), while
the truth values assigned by different functions in $\Sigma _{S}$ to $E(x)$
may differ if $E$ $\notin $ $\mathcal{E}_{S}\cup \mathcal{E}_{S}^{\bot }$.%
\footnote{%
Assumption A$_{4}$ can be stated unchanged whenever the standard
interpretation of QM is adopted instead of the SR interpretation. In this
case, however, for every $\xi $, $\sigma (\xi )$ is defined only on a subset
of rfs, not on the whole $\psi _{R}^{Q}$ (which requires a weakening of the
assumptions on $\sigma $ if one wants to recover this case within the
general perspective in Sec. 3.1). Furthermore, $\Sigma _{S}$ reduces to a
singleton. Indeed, for every interpretation $\xi $, a state $S=S(\xi )$
exists such that $\xi (x)\in S$. Then, $\sigma (\xi )$ is defined on a rf $%
E(x)$ iff $E\in \mathcal{E}_{S}\cup \mathcal{E}_{S}^{\bot }$ (Sec. 2.4), and
does not change if $\xi $ is substituted by an interpretation $\xi ^{\prime
} $ such that $\xi ^{\prime }(x)\in S$.}

Let us come now to the pragmatics of $\mathcal{L}_{Q}^{P}$. We introduce the
following assumption on $\mathcal{L}_{Q}^{P}$.

\smallskip

A$_{5}$. \textit{Let a mapping }$\xi $\textit{\ be given which interpretes
the variable }$x$\textit{\ in the rfs of }$\mathcal{L}_{Q}^{P}$\textit{\ on
a physical object in the state }$S$\textit{. A proof that the rf }$E(x)$%
\textit{\ is true (false) consists in performing one of the empirical
procedures mentioned in Sec. 2.6 and showing that }$E\in \mathcal{E}_{S}$%
\textit{\ (}$E\in \mathcal{E}_{S}^{\bot }$\textit{).}

\smallskip

Assumption A$_{5}$ is obviously suggested by the intended interpretation
discussed above. Taking into account A$_{1}$ and JR$_{1}$ in Sec. 3.1, it
implies the following statement.

\smallskip

P.\textit{\ Let }$E(x)$\textit{\ be a rf of }$\mathcal{L}_{Q}^{P}$\textit{,
let }$\xi $\textit{\ be an interpretation of the variable }$x$\textit{\ on a
physical object in the state }$S$\textit{, and let }$S_{E}$\textit{\ be
defined as in Sec. 2.2. Then,}

$\pi _{\sigma (\xi )}(\vdash E(x))=J$\textit{\ iff }$S\in $\textit{\ }$S_{E}$%
\textit{,}

$\pi _{\sigma (\xi )}(\vdash E(x))=U$\textit{\ iff }$S\notin $\textit{\ }$%
S_{E}$\textit{.}

\smallskip

The above result specifies $\pi _{\sigma (\xi )}$ on the set of all
elementary afs of $\mathcal{L}_{Q}^{P}$ and shows that it depends only on
the state $S$. Hence, we write $\pi _{S}$ in place of $\pi _{\sigma (\xi )}$
in the following (for the sake of brevity, we also agree to use the
intuitive statement ``the physical object $x$ is in the state $S$''
introduced in Sec. 2.1 in place of the more rigorous statement ``the
variable $x$ is interpreted on a physical object in the state $S$'').

Statement P provides the starting point for introducing a \textit{%
set-theoretical pragmatics} for $\mathcal{L}_{Q}^{P}$, as follows.

Firstly, we introduce a mapping

\smallskip

$f:\delta \in \psi _{A}^{Q}\longrightarrow \mathcal{S}_{\delta }\in \mathcal{%
P(S)}$

\smallskip

which associates a \textit{pragmatic extension }$\mathcal{S}_{\delta }$ with
every assertive formula $\delta \in \psi _{A}^{Q}$, defined by the following
recursive rules.

\smallskip

(i) \textit{For every }$E(x)\in \psi _{R}^{Q}$\textit{, }$f(\vdash
E(x))=S_{\vdash E(x)}=S_{E}$\textit{.}

(ii) \textit{For every }$\delta $\textit{\ }$\in \psi _{A}^{Q}$\textit{, }$%
f(N\delta )=S_{N\delta }=S_{\delta }^{\bot }$\textit{.}

(iii)\textit{\ For every }$\delta _{1}$\textit{, }$\delta _{2}\in \psi
_{A}^{Q}$\textit{, }$f(\delta _{1}K$ $\delta _{2})=\mathcal{S}_{\delta
_{1}K\delta _{2}}=\mathcal{S}_{\delta _{1}}\cap \mathcal{S}_{\delta _{2}}$.

(iv) \textit{For every }$\delta _{1}$\textit{, }$\delta _{2}\in \psi
_{A}^{Q} $\textit{, }$f(\delta _{1}A$\textit{\ }$\delta _{2})=S_{\delta
_{1}A\delta _{2}}=S_{\delta _{1}}\cup S_{\delta _{2}}$\textit{.}

\smallskip

Secondly, we rewrite statement P above substituting $\mathcal{S}_{\vdash
E(x)}$ to $\mathcal{S}_{E}$ in it.

\smallskip

P$^{\prime }$.\textit{\ Let }$\vdash E(x)$\textit{\ be an elementary af of }$%
\mathcal{L}_{Q}^{P}$\textit{\ and let }$x$ \textit{be in the state }$S$%
\textit{. Then,}

$\pi _{S}(\vdash E(x))=J$\textit{\ iff }$S\in $\textit{\ }$S_{\vdash E(x)}$%
\textit{,}

$\pi _{S}(\vdash E(x))=U$\textit{\ iff }$S\notin $\textit{\ }$S_{\vdash
E(x)} $\textit{.}

\smallskip

Thirdly, we note that statement P$^{\prime }$ defines the pragmatic
evaluation function $\pi _{S}$ on all elementary afs of $\mathcal{L}_{Q}^{P}$%
.

Finally, for every $S\in \mathcal{S}$, we extend $\pi _{S}$ from the set of
all elementary afs of $\mathcal{L}_{Q}^{P}$ to the set $\psi _{A}^{Q}$ of
all afs of $\mathcal{L}_{Q}^{P}$ bearing in mind JR$_{2}$ and JR$_{3}$ in
Sec. 3.1, hence introducing the following recursive rules.

\smallskip

(i) \textit{For every }$\delta $\textit{\ }$\in \psi _{A}^{Q}$\textit{, }$%
\pi _{S}(N\delta )=J$\textit{\ iff }$S\in S_{N\delta }=S_{\delta }^{\bot }$%
\textit{.}

(ii)\textit{\ For every }$\delta _{1}$\textit{, }$\delta _{2}\in \psi
_{A}^{Q}$\textit{, }$\pi _{S}(\delta _{1}K$\textit{\ }$\delta _{2})=J$%
\textit{\ iff }$S\in S_{\delta _{1}K\delta _{2}}=S_{\delta _{1}}\cap
S_{\delta _{2}}$\textit{.}

(iii) \textit{For every }$\delta _{1}$\textit{, }$\delta _{2}\in \psi
_{A}^{Q}$\textit{, }$\pi _{S}(\delta _{1}A$\textit{\ }$\delta _{2})=J$%
\textit{\ iff }$S\in S_{\delta _{1}A\delta _{2}}=S_{\delta _{1}}\cup
S_{\delta _{2}}$\textit{.}

\smallskip

The above procedure defines, for every $S\in \mathcal{S}$, a pragmatic
evaluation function

\smallskip

$\pi _{S}:\delta \in \psi _{A}^{Q}\longrightarrow \pi _{S}(\delta )\in
\{J,U\}$

\smallskip

which provides a set-theoretical pragmatics for $\mathcal{L}_{Q}^{P}$, as
stated.

\subsection{On the notion of justification in $\mathcal{L}_{Q}^{P}$}

The notion of justification introduced in Sec. 3.2 is basic in our approach
and must be clearly understood. So we devote this section to comments on it.

Whenever an elementary af $\vdash E(x)$ of $\mathcal{L}_{Q}^{P}$ is
considered, the notion of justification obviously coincides with the notion
of existence of an empirical proof of the truth of $E(x)$ because of
assumption A$_{5}$ and proposition P in Sec. 3.2, which fits in with JR$_{1}$
in Sec. 3.1.

Whenever molecular afs of $\mathcal{L}^{P}$ are considered, one can grasp
intuitively the meaning of the notion of justification for them by
considering simple instances. Indeed, let $E(x)$ be a rf and let $x$ be in
the state $S$. We get

\smallskip

$\pi _{S}(N\vdash E(x))=J$ iff $S\in \mathcal{S}_{E}^{\bot }$,

\smallskip

which means, shortly, that it is justified to assert that $E(x)$ cannot be
asserted iff MQ entails that the truth value of $E(x)$ is \textit{false} for
every $x$ in the state $S$. This result, of course, fits in with JR$_{2}$ in
Sec. 3.1.

Furthermore, let $E(x)$ and $F(x)$ be rfs, and let $x$ be in the state $S$.
We get

\smallskip

$\pi _{S}(\vdash E(x)K\vdash F(x))=J$ iff $S\in \mathcal{S}_{E}\cap \mathcal{%
S}_{F}$,

\smallskip

$\pi _{S}(\vdash E(x)A\vdash F(x))=J$ iff $S\in \mathcal{S}_{E}\cup \mathcal{%
S}_{F}$.

\smallskip

The first equality shows that asserting $E(x)$ and $F(x)$ conjointly is
justified iff both assertions are justified. The second equality shows that
asserting $E(x)$ or asserting $F(x)$ is justified iff one of these
assertions is justified. Both these results, of course, fit in with JR$_{3}$
in Sec. 3.1.

We add that

\smallskip

$\pi _{S}(\vdash E(x))=J$ implies $\pi _{S}(N\vdash E(x))=U$

\smallskip

and

\smallskip

$\pi _{S}(N\vdash E(x))=J$ implies $\pi _{S}(\vdash E(x))=U$

\smallskip

since $\mathcal{S}_{E}\cap \mathcal{S}_{E}^{\bot }=\emptyset $. Nevertheless,

\smallskip

$\pi _{S}(\vdash E(x))=U$ and $\pi _{S}(N\vdash E(x))=U$ iff $S\notin 
\mathcal{S}_{E}\cup \mathcal{S}_{E}^{\bot }$,

\smallskip

which shows that a \textit{tertium non datur }principle does not hold for
the pragmatic connective $N$ in $\mathcal{L}_{Q}^{P}$ (it has already been
proved in Ref. 27 that this principle does not hold in the general language $%
\mathcal{L}^{P}$).

It is also interesting to note that the justification values of different
elementary afs, say $\vdash E(x)$ and $\vdash F(x)$, must be different for
some state $S$, since $\mathcal{S}_{E}\neq \mathcal{S}_{F}$ if $E\neq F$
(Sec. 2.2), hence $\mathcal{S}_{\vdash E(x)}\neq \mathcal{S}_{\vdash F(x)}$.

Finally, we remind that the general theory of $\mathcal{L}^{P}$ associates
an assignment function $\sigma $ with a set $\Pi _{\sigma }$ of pragmatic
evaluation functions (Sec. 3.1), hence this also occurs within $\mathcal{L}%
_{Q}^{P}$. One may then wonder whether $\Pi _{\sigma }$ is necessarily
nonvoid and, if this is the case, whether it may contain more than one
pragmatic evaluation function. In order to answer these questions, let us
consider an interpretation $\xi $ of the variable $x$ that maps $x$ on a
physical object in the state $S$. Then, $\xi $ determines a unique
assignment function $\sigma (\xi )$ and a unique pragmatic evaluation
function associated with it, that we have denoted by $\pi _{S}$, for it
depends only on the state $S$. Since every assigment function in $\Sigma $
is induced by an interpretation $\xi $ because of A$_{4}$ in Sec. 3.2, this
proves that $\Pi _{\sigma }$ is necessarily nonvoid for every $\sigma \in
\Sigma $. Moreover, note that an interpretation $\xi ^{\prime }$ of $x$ may
exist within the SR interpretation of QM that maps $x$ on a physical object
in the state $S^{\prime }$, with $S^{\prime }\neq S$, yet such that $\sigma
(\xi ^{\prime })=\sigma (\xi )$. The pragmatic evaluation functions $\pi
_{S} $ and $\pi _{S^{\prime }}$ are then different, but they are both
associated with the assignment function $\sigma =\sigma (\xi )=\sigma (\xi
^{\prime })$, so that they both belong to $\Pi _{\sigma }$. Hence, $\Pi
_{\sigma }$ may contain many pragmatic evaluation functions.\footnote{%
Assumption A$_{5}$ in Sec. 3.2 can be stated unchanged if the standard
interpretation of QM is adopted instead of the SR interpretation. In this
case, however, it is impossible that a mapping $\xi ^{\prime }$ exists such
that $\xi ^{\prime }(x)\in S^{\prime }$, with $S\neq S^{\prime }$ and $%
\sigma (\xi )=\sigma (\xi ^{\prime })$, since $\sigma (\xi )$ and $\sigma
(\xi ^{\prime })$ are defined on different domains ($\mathcal{E}_{S}\cup 
\mathcal{E}_{S}^{\bot }$ and $\mathcal{E}_{S^{\prime }}\cup \mathcal{E}%
_{S^{\prime }}^{\bot }$, respectively). Hence, an assigment function $\sigma 
$ is associated with a unique state $S$, and $\Pi _{\sigma }$ reduces to the
singleton $\{\pi _{S}\}$.}

\subsection{Pragmatic validity and order in $\mathcal{L}_{Q}^{P}$}

Coming back to the general language $\mathcal{L}^{P}$, we remind that a
notion of pragmatic validity (invalidity) is introduced in it by means of
the following definition.

\smallskip

\textit{Let }$\delta \in \psi _{A}$\textit{. Then, }$\delta $\textit{\ is }%
pragmatically valid\textit{, or }p-valid\textit{\ (}pragmatically invalid%
\textit{, or }p-invalid\textit{) iff for every }$\sigma \in \Sigma $\textit{%
\ and }$\pi _{\sigma }\in \Pi _{\sigma }$\textit{, }$\pi _{\sigma }(\delta
)=J$\textit{\ (}$\pi _{\sigma }(\delta )=U$\textit{).}

\smallskip

By using the notions of justification in $\mathcal{L}_{Q}^{P}$, one can
translate the notion of p-validity (p-invalidity) within $\mathcal{L}%
_{Q}^{P} $ as follows.

\smallskip

\textit{Let }$\delta \in \psi _{A}^{Q}$\textit{. Then, }$\delta $\textit{\
is p-valid (p-invalid) iff, for every }$S\in S$\textit{, }$\pi _{S}(\delta
)=J$\textit{\ (}$\pi _{S}(\delta )=U$\textit{).}

\smallskip

The notion of p-validity (p-invalidity) can then be characterized as follows.

\smallskip

\textit{Let }$\delta \in \psi _{A}^{Q}$\textit{. Then, }$\delta $\textit{\
is p-valid (p-invalid) iff }$S_{\delta }=S$\textit{\ (}$S_{\delta
}=\emptyset $\textit{).}

\smallskip

The set of all p-valid afs plays in $\mathcal{L}_{Q}^{P}$ a role similar to
the role of tautologies in classical logic, and some afs in it can be
selected as axioms if one tries to construct a p-correct and p-complete
calculus for $\mathcal{L}_{Q}^{P}$. We will not deal, however, with this
topic in the present paper.

Furthermore, let us observe that a binary relation can be introduced in the
general language $\mathcal{L}^{P}$ by means of the following definition.

\smallskip

\textit{For every }$\delta _{1}$\textit{, }$\delta _{2}\in \psi _{A}$\textit{%
, }$\delta _{1}\prec $\textit{\ }$\delta _{2}$\textit{\ iff a proof exists
that }$\delta _{2}$\textit{\ is justified whenever }$\delta _{1}$\textit{\
is justified (equivalently, }$\delta _{1}\prec \delta _{2}$\textit{\ iff }$%
\delta _{1}C\delta _{2}$\textit{\ is justified}).

\smallskip

The set-theoretical pragmatics introduced in Sec. 3.2 allows one to
translate the above definition in $\mathcal{L}_{Q}^{P}$ as follows.

\smallskip

\textit{For every }$\delta _{1}$\textit{, }$\delta _{2}\in \psi _{A}^{Q}$%
\textit{, }$\delta _{1}\prec $\textit{\ }$\delta _{2}$\textit{\ iff for
every }$S\in S$\textit{, }$\pi _{S}(\delta _{1})=J$\textit{\ implies }$\pi
_{S}(\delta _{2})=J$\textit{.}

\smallskip

The binary relation $\prec $ can then be characterized as follows.

\smallskip

\textit{For every }$\delta _{1}$\textit{, }$\delta _{2}\in \psi _{A}^{Q}$%
\textit{, }$\delta _{1}\prec $\textit{\ }$\delta _{2}$\textit{\ iff }$%
S_{\delta _{1}}\subset S_{\delta _{2}}$\textit{.}

\smallskip

The relation $\prec $ is obviously a pre-order relation on $\psi _{A}^{Q}$,
hence it induces canonically an equivalence relation $\approx $ on $\psi
_{A}^{Q}$, defined as follows.

\smallskip

\textit{For every }$\delta _{1}$\textit{, }$\delta _{2}\in \psi _{A}^{Q}$%
\textit{, }$\delta _{1}\approx $\textit{\ }$\delta _{2}$\textit{\ iff }$%
\delta _{1}\prec $\textit{\ }$\delta _{2}$\textit{\ and }$\delta _{2}\prec $%
\textit{\ }$\delta _{1}$\textit{.}

\smallskip

The equivalence relation $\approx $ can then be characterized as
follows.\smallskip

\smallskip

\textit{For every }$\delta _{1}$\textit{, }$\delta _{2}\in \psi _{A}^{Q}$%
\textit{, }$\delta _{1}\approx $\textit{\ }$\delta _{2}$\textit{\ iff }$%
S_{\delta _{1}}=S_{\delta _{2}}$\textit{.}

\subsection{Decidability versus justifiability in $\mathcal{L}_{Q}^{P}$}

We have commented rather extensively in Sec. 3.3 on the notion of
justification formalized in $\mathcal{L}_{Q}^{P}$, for every $S\in \mathcal{S%
}$, by the pragmatic evaluation function $\pi _{S}$. It must still be noted,
however, that the definition of $\pi _{S}$ on all afs in $\psi _{A}^{Q}$
does not grant that an empirical procedure of proof exists which allows one
to establish, for every $S\in \mathcal{S}$, the justification value of every
af of $\mathcal{L}_{Q}^{P}$. In order to understand how this may occur, note
that the notion of empirical proof is defined by A$_{5}$ for atomic rfs of $%
\mathcal{L}_{Q}^{P}$ and makes explicit reference, for every $E(x)\in \psi
_{R}^{Q}$, to the closed subset $\mathcal{S}_{E}\in \mathcal{L(S)}$
associated with $E$ by the function $\rho $ introduced in Sec. 2.2. Basing
on this notion, the justification value $\pi _{S}(\vdash E(x))$ of an
elementary af $\vdash E(x)\in \psi _{A}^{Q}$ can be determined by means of
the same empirical procedure, making reference to the closed subset $%
\mathcal{S}_{\vdash E(x)}=\mathcal{S}_{E}$ associated to $\vdash E(x)$ by
the function $f$ (Sec. 3.2). Yet, whenever $\pi _{S}$ is recursively defined
on the whole $\psi _{A}^{Q}$, new subsets of states are introduced (as $%
\mathcal{S}_{\delta _{1}}\cup \mathcal{S}_{\delta _{2}}$) which do not
necessarily belong to $\mathcal{L(S)}$. If an af $\delta $ is associated by $%
f$ with a subset that does not belong to $\mathcal{L(S)}$, no empirical
procedure exists in QM which allows one to determine the justification value 
$\pi _{S}(\delta )$.

We are thus led to introduce the subset $\psi _{AD}^{Q}$ $\subset \psi
_{A}^{Q}$ of all \textit{pragmatically decidable}, or \textit{p-decidable},
afs of $\mathcal{L}_{Q}^{P}$. An af $\delta $ of $\mathcal{L}_{Q}^{P}$ is
p-decidable iff an empirical procedure of proof exists which allows one to
establish whether $\delta $ is justified or unjustified, whatever the state $%
S$ of $x$ may be.

Because of the remark above, the subset of all p-decidable afs of $\mathcal{L%
}_{Q}^{P}$ can be characterized as follows.

\smallskip

$\psi _{AD}^{Q}=\{\delta \in \psi _{A}^{Q}\mid $ $\mathcal{S}_{\delta }\in 
\mathcal{L(S)}\}$.

\smallskip

Let us discuss some criteria for establishing whether a given af $\delta \in 
$ $\psi _{A}^{Q}$ belongs to $\psi _{AD}^{Q}$.

\smallskip

C$_{1}$. \textit{All elementary afs of }$\psi _{A}^{Q}$\textit{\ belong to }$%
\psi _{AD}^{Q}$\textit{.}

\smallskip

C$_{2}$. \textit{If }$\delta \in $\textit{\ }$\psi _{AD}^{Q}$\textit{, then }%
$N\delta \in $\textit{\ }$\psi _{AD}^{Q}$\textit{\ }

\smallskip

Indeed, $S_{\delta }\in \mathcal{L(S)}$\ implies $S_{\delta }^{\bot }\in 
\mathcal{L(S)}$.

\smallskip

C$_{3}$. \textit{If }$\delta _{1}$\textit{, }$\delta _{2}\in $\textit{\ }$%
\psi _{AD}^{Q}$\textit{, then }$\delta _{1}K$\textit{\ }$\delta _{2}\in $%
\textit{\ }$\psi _{AD}^{Q}$

\smallskip

Indeed, $S_{\delta _{1}}\in \mathcal{L(S)}$\ and $S_{\delta _{2}}\in 
\mathcal{L(S)}$\ imply $S_{\delta _{1}}\cap S_{\delta _{2}}\in \mathcal{L(S)}
$, since $S_{\delta _{1}}\cap S_{\delta _{2}}=S_{\delta _{1}}\Cap S_{\delta
_{2}}$\ because of known properties of the lattice $(\mathcal{L(S)},\subset
) $ (Sec. 2.2).

\smallskip

C$_{4}$. \textit{If }$\delta _{1}$\textit{, }$\delta _{2}\in $\textit{\ }$%
\psi _{AD}^{Q}$\textit{, then }$\delta _{1}A$\textit{\ }$\delta _{2}$\textit{%
\ may belong or not to }$\psi _{AD}^{Q}$\textit{. To be precise, it belongs
to }$\psi _{AD}^{Q}$\textit{\ iff }$S_{\delta _{1}}\subset S_{\delta _{2}}$%
\textit{\ or }$S_{\delta _{2}}\subset S_{\delta _{1}}$

\smallskip

Indeed, $S_{\delta _{1}}\cup S_{\delta _{2}}\in \mathcal{L(S)}$\ or,
equivalently, $S_{\delta _{1}}\cup S_{\delta _{2}}=S_{\delta _{1}}\Cup
S_{\delta _{2}}$, iff one of the conditions in C$_{4}$ is satisfied.

It is apparent from criteria C$_{2}$ and C$_{3}$ that $\psi _{AD}^{Q}$ is
closed with respect to the pragmatic connectives $N$ and $K$, in the sense
that $\delta \in \psi _{AD}^{Q}$ implies $N\delta \in \psi _{AD}^{Q}$, and $%
\delta _{1}$, $\delta _{2}\in \psi _{AD}^{Q}$ implies $\delta _{1}K\delta
_{2}\in \psi _{AD}^{Q}$. On the contrary, $\psi _{AD}^{Q}$ is not closed
with respect to $A$, since it may occur that $\delta _{1}A$ $\delta
_{2}\notin \psi _{AD}^{Q}$ even if $\delta _{1}$, $\delta _{2}\in \psi
_{AD}^{Q}$. In order to obtain a closed subset of afs of $\mathcal{L}%
_{Q}^{P} $, one can consider the set

\smallskip

$\phi _{AD}^{Q}=\{\delta \in \psi _{A}^{Q}\mid $ the pragmatic connective $A$
does not occur in $\delta \}$.

\smallskip

The set $\phi _{AD}^{Q}$ obviously contains all elementary afs of $\mathcal{L%
}_{Q}^{P}$, plus all afs of $\psi _{A}^{Q}$ in which only the pragmatic
connectives $N$ and $K$ occur. We can thus consider a sublanguage of $%
\mathcal{L}_{Q}^{P}$ whose set of afs reduces to $\phi _{AD}^{Q}$. This new
language is relevant since all its afs are p-decidable, hence we call it%
\textit{\ the} \textit{p-decidable sublanguage} of $\mathcal{L}_{Q}^{P}$ and
denote it by $\mathcal{L}_{QD}^{P}$.

\subsection{The p-decidable sublanguage $\mathcal{L}_{QD}^{P}$}

As we have anticipated in the Introduction, we aim to show in this paper
that the sublanguage $\mathcal{L}_{QD}^{P}$ has the structure of a physical
QL, hence it provides a new pragmatic interpretation of this relevant
physical structure. However, this interpretation will be more satisfactory
from an intuitive viewpoint if we endow $\mathcal{L}_{QD}^{P}$ with some
further derived pragmatic connectives which can be made to correspond with
connectives of physical QL. To this end, we introduce the following
definitions.

\smallskip

D$_{1}$. \textit{We call }quantum pragmatic disjunction \textit{the
connective }$A_{Q}$\textit{\ defined as follows.}

\textit{For every }$\delta _{1}$\textit{, }$\delta _{2}\in $\textit{\ }$\phi
_{AD}^{Q}$\textit{, }$\delta _{1}A_{Q}\delta _{2}=N((N\delta _{1})K(N\delta
_{2}))$\textit{.}

\smallskip

D$_{2}$. \textit{We call }quantum pragmatic implication\textit{\ the
connective }$I_{Q}$\textit{\ defined as follows.}

\textit{For every }$\delta _{1}$\textit{, }$\delta _{2}\in $\textit{\ }$\phi
_{AD}^{Q}$\textit{, }$\delta _{1}I_{Q}\delta _{2}=(N\delta _{1})A_{Q}(\delta
_{1}K\delta _{2})$\textit{.}

\smallskip

Let us discuss the justification rules which hold for afs in which the new
connectives $A_{Q}$ and $I_{Q}$ occur.

By using the function $f$ introduced in Sec. 3.2 we get (since the
set-theoretical operation $\cap $ coincides with the lattice operation $\Cap 
$ in $(\mathcal{L(S)},\subset )$, see Sec. 2.2),

\smallskip

$\mathcal{S}_{\delta _{1}A_{Q}\delta _{2}}=\mathcal{S}_{(N\delta
_{1})K(N\delta _{2})}^{\bot }=(\mathcal{S}_{N\delta _{1}}\cap \mathcal{S}%
_{N\delta _{2}})^{\bot }=(\mathcal{S}_{\delta _{1}}^{\bot }\Cap \mathcal{S}%
_{\delta _{2}}^{\bot })^{\bot }=(\mathcal{S}_{\delta _{1}}\Cup \mathcal{S}%
_{\delta _{2}})$.

\smallskip

Hence, for every $S\in \mathcal{S}$,

\smallskip

$\pi _{S}(\delta _{1}A_{Q}\delta _{2})=J$ iff $S\in \mathcal{S}_{\delta
_{1}}\Cup \mathcal{S}_{\delta _{2}}$.

\smallskip

Let us come to the quantum pragmatic implication $I_{Q}$. By using the
definition of $A_{Q}$, one gets

\smallskip

$\delta _{1}I_{Q}\delta _{2}=N((NN\delta _{1})K(N(\delta _{1}K\delta _{2}))$.

\smallskip

By using the function $f$ and the above result about $A_{Q}$, one then gets

\smallskip

$\mathcal{S}_{\delta _{1}I_{Q}\delta _{2}}=\mathcal{S}_{N\delta _{1}}\Cup 
\mathcal{S}_{\delta _{1}K\delta _{2}}=\mathcal{S}_{\delta _{1}}^{\bot }\Cup (%
\mathcal{S}_{\delta _{1}}\Cap \mathcal{S}_{\delta _{2}})$.

\smallskip

It follows that, for every $S\in \mathcal{S}$,

\smallskip

$\pi _{S}(\delta _{1}I_{Q}\delta _{2})=J$ iff $S\in \mathcal{S}_{\delta
_{1}}^{\bot }\Cup (\mathcal{S}_{\delta _{1}}\Cap \mathcal{S}_{\delta _{2}})$.

\smallskip

Let us observe now that $\mathcal{L}_{QD}^{P}$ obviously inherits the
notions of p-validity and order defined in $\mathcal{L}_{Q}^{P}$ (Sec. 3.4).
Hence, we can illustrate the role of the connective $I_{Q}$ within $\mathcal{%
L}_{QD}^{P}$ by means of the following \textit{pragmatic deduction lemma}.

\smallskip

PDL.\textit{\ Let }$\delta _{1}$\textit{, }$\delta _{2}\in $\textit{\ }$\phi
_{AD}^{Q}$\textit{. Then, }$\delta _{1}\prec $\textit{\ }$\delta _{2}$%
\textit{\ iff for every }$S\in S$\textit{, }$\pi _{S}(\delta _{1}I_{Q}\delta
_{2})=J$\textit{\ (equivalently, iff }$\delta _{1}I_{Q}\delta _{2}$\textit{\
is p-valid).}

\smallskip

Proof. The following sequence of equivalences holds.

For every $S\in \mathcal{S}$, $\pi _{S}(\delta _{1}I_{Q}\delta _{2})=J$ iff
for every $S\in \mathcal{S}$, $S\in \mathcal{S}_{\delta _{1}}^{\bot }\Cup (%
\mathcal{S}_{\delta _{1}}\Cap \mathcal{S}_{\delta _{2}})$ iff $\mathcal{S}%
_{\delta _{1}}^{\bot }\Cup (\mathcal{S}_{\delta _{1}}\Cap \mathcal{S}%
_{\delta _{2}})=\mathcal{S}$ iff $\mathcal{S}_{\delta _{1}}\Cap \mathcal{S}%
_{\delta _{2}}=$ $\mathcal{S}_{\delta _{1}}$ iff $\mathcal{S}_{\delta
_{1}}\subset \mathcal{S}_{\delta _{2}}$ iff $\delta _{1}\prec $ $\delta _{2}$%
.$\blacksquare \smallskip $\smallskip

PDL shows that the quantum pragmatic implication $I_{Q}$ plays within $%
\mathcal{L}_{QD}^{P}$ a role similar to the role of material implication in
classical logic.

\subsection{Interpreting QL onto $\mathcal{L}_{QD}^{P}$}

In order to show that the physical QL $(\mathcal{E},\prec )$ introduced in
Sec. 2.2 can be interpreted into $\mathcal{L}_{QD}^{P}$, a further
preliminary step is needed. To be precise, let us make reference to the
preorder introduced on $\psi _{A}^{Q}$ in Sec. 3.4 and consider the
pre-ordered set $(\phi _{AD}^{Q},\prec )$ of all afs of $\mathcal{L}%
_{QD}^{P} $. Furthermore, let us denote by $\approx $ (by abuse of language)
the restriction of the equivalence relation introduced on $\psi _{A}^{Q}$ in
Sec. 3.4 to $\phi _{AD}^{Q}$, and let us denote by $\prec $ (again by abuse
of language) the partial order induced on $\phi _{AD}^{Q}/\approx $ by the
preorder defined on $\phi _{AD}^{Q}$. Then, let us show that $(\phi
_{AD}^{Q}/\approx ,\prec )$ is order isomorphic to $(\mathcal{L(S)},\subset
) $.

Let us consider the mapping

\smallskip

$f_{\approx }:[\delta ]_{\approx }\in \psi _{AD}^{Q}/\approx
\;\longrightarrow $\ $\mathcal{S}_{\delta }\in \mathcal{L(S)}$.

\smallskip

This mapping is obviously well defined because of the characterization of $%
\approx $ in Sec. 3.4. Furthermore, the following statements hold.

\smallskip

(i) \textit{For every }$\delta \in \phi _{AD}^{Q}$\textit{, one and only one
elementary af }$\vdash E(x)$\textit{\ exists such that }$\vdash E(x)\in
\lbrack \delta ]_{\approx }$.

(ii) \textit{The mapping }$f_{\approx }$\textit{\ is bijective.}

(iii) \textit{For every }$\delta _{1}$\textit{, }$\delta _{2}\in $\textit{\ }%
$\phi _{AD}^{Q}$\textit{, }$[\delta _{1}]_{\approx }\prec \lbrack \delta
_{2}]_{\approx }$\textit{\ iff }$S_{\delta _{1}}\subset S_{\delta _{2}}$%
\textit{.}

\smallskip

Let us prove (i). Consider $[\delta ]_{\approx }$. Since $\mathcal{S}%
_{\delta }\in \mathcal{L(S)}$ and $\rho $ is bijective (Sec. 2.2), a
property $E\in \mathcal{E}$ exists such that $E=\rho ^{-1}(\mathcal{S}%
_{\delta })$, hence $\mathcal{S}_{\delta }=\mathcal{S}_{E}$. It follows that 
$[\delta ]_{\approx }$ contains the af $\vdash E(x)$, for $\mathcal{S}%
_{\vdash E(x)}=\mathcal{S}_{E}$ (Sec. 3.2). Moreover, $[\delta ]_{\approx }$
does not contain any further elementary af. Indeed, let $\vdash F(x)$ be an
elementary af of $\phi _{AD}^{Q}$ with $E\neq F$: then, $\mathcal{S}_{E}\neq 
\mathcal{S}_{F}$, hence $\mathcal{S}_{\vdash E(x)}\neq \mathcal{S}_{\vdash
F(x)}$, which implies $\vdash F(x)\notin \lbrack \delta ]_{\approx }$. Thus,
statement (i) is proved.

The proofs of statements (ii) and (iii) are then immediate. Indeed,
statement (ii) follows from (i) and from the definition of $f_{\approx }$,
while statement (iii) follows from (ii) and from the definition of $\prec $
on $\phi _{AD}^{Q}/\approx $.

Because of (ii) and (iii), the poset $(\phi _{AD}^{Q}/\approx ,\prec )$ is
order-isomorphic to $(\mathcal{L(S)},\subset )$, as stated.

Let us come now to physical QL. We have seen in Sec. 2.2 that $(\mathcal{L(S)%
},\subset )$ is order-isomorphic to $(\mathcal{E},\prec )$. We can then
conclude that $(\mathcal{E},\prec )$ is order-isomorphic to $(\phi
_{AD}^{Q}/\approx ,\prec )$, which provides the desired interpretation of a
physical QL into $\mathcal{L}_{QD}^{P}$.

Let us comment briefly on the pragmatic interpretation of physical QL
provided above.

Firstly, we note that our interpretation maps $\mathcal{E}$ on the quotient
set $\phi _{AD}^{Q}/\approx $, not onto $\phi _{AD}^{Q}$. Yet, the set of
the (well formed) formulas of the lattice $(\mathcal{E},^{\bot },\Cap ,\Cup
) $ can be mapped bijectively onto $\phi _{AD}^{Q}$ by means of the mapping
induced by the following formal correspondence.

\smallskip

(i) $E\in \mathcal{E}$ $\longleftrightarrow \vdash E(x)\in \phi _{AD}^{Q}$.

(ii) $^{\bot }\longleftrightarrow N$

(iii) $\Cap \longleftrightarrow K$

(iv) $\Cup \longleftrightarrow A_{Q}$.

\smallskip

Thus, the formal language of QL, for which the lattice $(\mathcal{L(S)}%
,\subset )$ can be considered as an \textit{algebraic semantics},$^{(3)}$
can be substituted by the language $\mathcal{L}_{QD}^{P}$, for which $(%
\mathcal{L(S)},\subset )$ can be considered as an \textit{algebraic
pragmatics} (by the way, we also note that the above correspondence makes $%
I_{Q}$ correspond to a \textit{Sasaki hook}, the role of which is well known
in QL). This reinterpretation is relevant from a philosophical viewpoint,
since it avoids all problems following from the standard concept of quantum
truth (Sec. 2.4) considering physical QL as formalizing properties of a
quantum concept of justification rather than a quantum concept of truth.
This makes physical QL consistent also with the classical concept of truth
adopted with the SR interpretation of QM (Sec. 2.5). Furthermore, as we have
already observed in the Introduction, it places physical QL within a general 
\textit{integrated perspective}, according to which non-Tarskian theories of
truth can be integrated with Tarski's theory by reinterpreting them as
theories of metalinguistic concepts that are different from truth (in the
case of physical QL, the concept of \textit{empirical justification} in QM).

Secondly, we observe that our interpretation has some consequences that are
intuitively satisfactory. For instance, for every state $S\in \mathcal{S}$,
it attributes a justification value to every af in $\phi _{AD}^{Q}$, while
it is well known that there are formulas in physical QL which have no truth
value according to the standard interpretation of QL (Sec. 2.4).

\subsection{Some remarks on a possible calculus for $\mathcal{L}_{QD}^{P}$}

One may obviously wonder whether a calculus can be given for the language $%
\mathcal{L}_{QD}^{P}$ which is \textit{pragmatically correct} (\textit{%
p-correct}) and \textit{pragmatically complete} (\textit{p-complete}). This
is not a difficult task if we limit ourselves to the general lattice
structure of $(\phi _{AD}^{Q}/\approx ,\prec )$. Indeed, a set of axioms
and/or inference rules which endow $\phi _{AD}^{Q}/\approx $ of the
structure of orthomodular lattice can be easily obtained by using the formal
correspondence introduced in Sec. 3.7, since this correspondence allows one
to translate the axioms and/or inference rules that are usually stated in
order to provide a calculus for orthomodular QL into $\phi _{AD}^{Q}$ (of
course, all the afs produced by this translation are p-valid afs of $%
\mathcal{L}_{QD}^{P}$). Here is a sample set of axioms of this kind (where,
of course, $\delta $, $\delta _{1}$, $\delta _{2}$ and $\delta _{3}$ are afs
of $\phi _{AD}^{Q}$) obtained by translating a set of rules provided by
Dalla Chiara and Giuntini.$^{(32)}$

\smallskip

A$_{1}$. $\delta I_{Q}\delta $.

\smallskip

A$_{2}$. $($ $\delta _{1}K$ $\delta _{2})I_{Q}\delta _{1}$.

\smallskip

A$_{3}$. $($ $\delta _{1}K$ $\delta _{2})I_{Q}\delta _{2}$.

\smallskip

A$_{4}$. $\delta I_{Q}(NN\delta )$.

\smallskip

A$_{5}$. $(NN\delta )I_{Q}\delta $.

\smallskip

A$_{6}$. $((\delta _{1}I_{Q}\delta _{2})K(\delta _{1}I_{Q}\delta
_{3}))I_{Q}(\delta _{1}I_{Q}(\delta _{2}K\delta _{3}))$.

\smallskip

A$_{7}$. $((\delta _{1}I_{Q}\delta _{2})K(\delta _{2}I_{Q}\delta
_{3}))I_{Q}(\delta _{1}I_{Q}\delta _{3})$.

\smallskip

A$_{8}$. $(\delta _{1}I_{Q}\delta _{2})I_{Q}((N\delta _{2})I_{Q}(N\delta
_{1}))$.

\smallskip

A$_{9}$. $(\delta _{1}I_{Q}\delta _{2})I_{Q}(\delta _{2}I_{Q}(\delta
_{1}A_{Q}((N\delta _{1})K\delta _{2})))$.

\smallskip

However, in order to obtain physical QL one needs a number of further
axioms, since the structure of $(\mathcal{L(H)},\subset )$ must be recovered
(Sec. 2.2). Providing a complete calculus for such a structure is a much
more complicate task, which must take into account a number of mathematical
results in lattice theory (in particular, Soler's theorem$^{(33)}$).
Therefore we will not discuss this problem in the present paper.

\bigskip

\bigskip

\bigskip

\bigskip

\bigskip \bigskip

\bigskip \textbf{ACKNOWLEDGEMENT}

\bigskip The author is greatly indebted to Carlo Dalla Pozza, Jaroslaw
Pykacz and Sandro Sozzo for reading the manuscript and providing many useful
suggestions.

\end{document}